\documentclass{aa}

\usepackage{graphics}

\begin{document}
  
  \thesaurus{08 (08.22.3; 08.09.2 HD 101947)}
\thesaurus{06         
          (08.09.2  V810 Centauri;   
           08.19.3;                  
           08.15.1;                  
           08.05.3;                  
           10.15.1  Stock 14;        
           03.20.4)                  
             }

  \title{The pulsating yellow supergiant V810~Centauri
           \thanks{Based on observations collected at the Swiss 40~cm and 70~cm and at the Danish 1.54~m telescopes,
	          at the European Southern Observatory (La Silla, Chile)}
              \thanks{ The photometric data are only available in electronic form at the CDS via anonymous ftp 
	               to cdsarc.u.strasbg.fr (130.79.128.5) or via http://cdsweb.u-strasbg.fr/Abstract.html}
                     }

   \author{F. Kienzle \inst{1} \and G. Burki \inst{1} \and M. Burnet \inst{1} \and G. Meynet \inst{1}}

   \institute{Observatoire de Genève,
              Ch. des Maillettes 51,  
              CH-1290 Sauverny,
              Switzerland
	     }

   \offprints{F. Kienzle}
   \mail{Francesco.Kienzle@obs.unige.ch}

   \date{Received / Accepted}

   \maketitle
   \markboth{F. Kienzle et al.: The pulsating yellow supergiant V810 Centauri}
            {F. Kienzle et al.: The pulsating yellow supergiant V810 Centauri}


\begin{abstract}

The F8\,Ia supergiant \object{V810~Centauri} is part of a long-term high-precision photometric monitoring
program on long period variables started twenty years ago. Time series analysis of this unique set of 500 
data points, spanning almost fifteen years in the homogeneous Geneva photometric system, is presented. 
Cluster membership, physical parameters and evolutionary status of the star are reinvestigated. Radial 
velocity data do not support the cluster membership to \object{Stock~14}. Ultraviolet and optical spectrophotometry
is combined with optical and infrared photometry to evaluate the physical parameters of the yellow
supergiant ($T_{\rm eff}$ = 5970 K, $M_{\rm bol}$ = -8.5, $R$ = 420 R$_{\odot}$) and of its B0\,III 
companion. From theoretical stellar evolutionary tracks, an initial mass of $\sim$25 M$_{\odot}$ is estimated
for \object{V810~Cen}, which is actually at the end of its first redward evolution. 

\object{V810~Cen} is a multi-periodic small amplitude variable star, whose amplitudes are variable with time. The
period of the main mode, $\sim$156~d, is in agreement with the Period--Luminosity--Colour relation
for supergiants. This mode is most probably the fundamental radial one. 
According to the theoretical pulsation periods for the radial modes, calculated from a linear non-adiabatic analysis,
the period of the observed second mode, $\sim$107~d, is much too long to correspond to the first radial 
overtone . Thus, this second mode could be a non-radial p-mode.
Other transient periods are observed, in particular at $\sim$187~d. The length of this period suggests a non-radial 
g-mode. Then, the complex variability of \object{V810~Cen} could be due to a mixing of unstable radial and non-radial
p- and g-modes.

   \keywords{Stars: individual : V810 Centauri -- Stars : supergiants -- Stars : oscillations --
   Stars : evolution -- Open clusters : individual : Stock 14 -- Techniques : photometry}
   \end{abstract}

\section{Introduction}

\object{V810~Centauri} (=\object{HD~101947} =\object{HR~4511}, hereafter \object{V810~Cen}) is a yellow supergiant
at low galactic latitude (l=295.18, b=-0.64) in the direction of Carina spiral arm.

The light variability was discovered by Fernie (\cite{fernie}) as a result of
cepheid-like supergiants photometric survey. Eichendorf \& Reipurth (\cite{eichendorf1}) found a
low amplitude variation (0.12 mag in $V$) with a ``period'' of 125 days (but only two
successive minima were observed). Dean (\cite{dean}) confirmed the variability of
\object{V810~Cen}, but not the 125~d period. In a preliminary analysis of Geneva long-term 
photometric monitoring, Burki (\cite{burki2}) obtained two radial modes at 153 and 
104 days with low amplitudes (0.07 and 0.04 mag respectively). However,
the high residuals suggested that the amplitudes were variable and/or that 
additional frequencies may be present.

The pulsation interpretation of the variability was controversial.
Indeed, Bidelman et al. (\cite{bidelman}) suspected the star to be a spectroscopic binary and
this was confirmed by van Genderen (\cite{vangenderen1}) who proposed a B spectral type companion 
to account for the UV excess of the G0 Ia star. Therefore, the light variation may
come from the companion variability rather than the G0 star. This issue was settled
with IUE spectra from Parsons (\cite{parsons2}). From continuum flux fitting and photometric data
he found that the G supergiant should be 3.2 mag brighter (in the $V$ band) than its
B companion, thus comforting the cepheid-like pulsation hypothesis for \object{V810~Cen}.
However, the spectral type of the B companion was still ambiguous: C~IV and Si~IV absorption
lines lead to a B0-B1~Iab-Ib star while continuum flux fitting requires a B0.5-B1~III
star (see Section~3). Since the G star luminosity relies on its companion bolometric magnitude
and their magnitude difference, the variable star luminosity is also ambiguous.

The perspective of using \object{V810~Cen} as a prime candidate to understand the variability 
of the massive stars was strengthened by the open cluster membership. Moffat \& Vogt (\cite{moffat})
suggested that \object{V810~Cen} is a member of the open cluster \object{Stock~14} for which Peterson \& FitzGerald
(\cite{peterson}) derived $<E_{B-V}>=0.26$ and $V_{0}-\mathrm{M_{V}}=12.14$. But the intrinsic 
luminosity, derived from the cluster distance, is 0.5 mag fainter when compared to
the spectroscopic estimates from IUE data.

In the present paper we re-investigate various aspects of this long-period variable star. Section~2 
is devoted to the discussion of the membership to the open cluster \object{Stock~14} based on the cluster members
radial velocity. IUE final archive and visual spectrophotometry are used in Section~4 
to constrain the luminosity and effective temperature of the G0 supergiant. The variability of \object{V810~Cen} is
discussed thoroughly on the basis of the Geneva photometric data, collected over 15 years (Sections 5 to 8).
Finally, a discussion of the pulsation modes is given in Section~9.

\section{V810~Cen and Stock~14}

Moffat \& Vogt (\cite{moffat}) postulated that \object{V810~Cen} is a member of the open cluster
\object{Stock~14} (\object{C 1141-622}). This would strongly constrain the physical
parameters of the supergiant. Unfortunately, as we shall see hereafter, this is
probably not the case.

With the UBV photometry from Moffat \& Vogt (\cite{moffat}), Turner (\cite{turner}) and
Peterson \& FitzGerald (\cite{peterson}),
collected by Mermilliod (\cite{mermillod}) in his Cluster Database, and using the stellar evolutionary 
tracks from Schaller et al. (\cite{schaller2}), the following parameters can be derived for
\object{Stock~14} : 
$E(B-V)=0.26$, distance= 2.63 kpc, age= $1.4 \cdot 10^{7}$ yr. These values are consistent 
with those found by Moffat \& Vogt (\cite{moffat}), FitzGerald \& Miller (\cite{fitzgerald}), Lynga (\cite{lynga}) and
Peterson \& FitzGerald (\cite{peterson}).

Assuming cluster membership, the absolute magnitude of \object{V810~Cen} (both components together)
would be $\mathrm{M_{V}}=-7.88$, with an UBV intrinsic color of $(B-V)_{0}=0.526$.
According to the analysis of van Genderen (\cite{vangenderen2}), the blue component has an absolute magnitude of
$\mathrm{M_{V}}=-4.25$ and an intrinsic color index of $(B-V)_{0}=-0.28$ (O9-B0 type star), thus the red 
supergiant would have $\mathrm{M_{V}}=-7.84$ and $(B-V)_{0}=0.56$, corresponding to a G0Ia type star.
With these values, the blue component would be close to the cluster turnoff of the main sequence, 
whereas the red supergiant remains close to the original position of the double system,
i.e. close to the blue loop of the evolutionary track, in the core helium burning phase.

As we shall see from the analysis of the radial velocity data, it is probable that
\object{V810~Cen} is not a member of \object{Stock~14}. The radial velocities of 4 bright
B-type members of this cluster are known: -8 km/s (1 measurement  with an
uncertainty of at least 5 km/s) for \object{HD~101994} (Buscombe \& Kennedy, ~\cite{buscombe}),
$-6.4 \pm 1.9$ km/s for the eclipsing SB2 system \object{V346 Cen}, \object{HD~101897}
(Hernandez \& Sahade (\cite{hernandez}), $-3 \pm 12$ km/s (4 measurements) for \object{HD~101964}
(Feast et al., \cite{feast1}) and $-10 \pm 21$ km/s (5 measurements, the velocity
could be variable) for \object{HD~101838} (Feast \& Thackeray, \cite{feast2}). A mean velocity of
$-6 \pm 2$ km/s can be adopted for the cluster in agreement with the estimation
of Lynga (\cite{lynga}). This value relies strongly on the $\gamma$ velocity of the
SB2 system \object{V346 Cen} from 44 $V_{r}$ measurements.

\object{V810~Cen} has 90 radial velocity data points between HJD 2\,444\,621 
and 2\,449\,915 from \textsc{Coravel} spectrophotometer attached to the 1.54~m Danish 
telescope in La Silla. The uncertainty on each measurement is about 0.4 km/s, whereas
the dispersion of the data is much larger, due to the complex pulsation of \object{V810~Cen},
which is the only component of the system measured with \textsc{Coravel}. The mean velocity is 
$-16.7\pm 2.6$ (s.d.). For further calculation,  the annual means of the velocity
will be used. The values of 13 annual means range from -13.2 km/s to -20.8 km/s, 
with standard deviations between 0.5 and 3.5 km/s.

Older measurements of the radial velocity of \object{V810~Cen} are: +11.7 km/s in 1908 
(2 measurements) and +6.9 km/s in 1911 (1 measurement) from Campbell \& Moore (\cite{campbell}), and
-17.4 km/s in 1946 (10 measurements), -14.4 km/s (3 measurements) in 1947 and -10.8 km/s
(2 measurements) in 1959 from Bidelman et al. (\cite{bidelman}).

\begin{figure}[thb]
  \resizebox{\hsize}{!}{\includegraphics{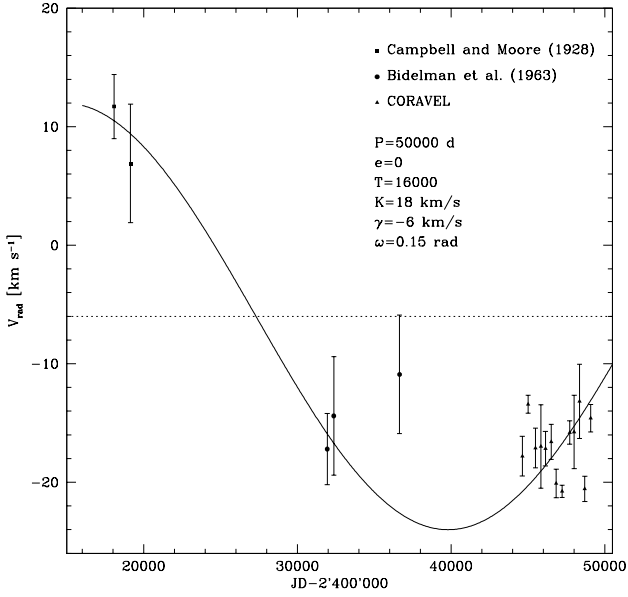}}
  \caption{Radial velocities from 1908 to 1995 collected from various sources.
    Only mean values are shown. The important scatter in the last set of data is
    due to V810~Cen pulsation. The fitted orbit (full curve) would agree with the
    1908-1928 data but is physically unlikely (see text).}
  \label{fig:vrad}
\end{figure}

If all the radial velocity data are taken into account, the possible orbits with
a $\gamma$ velocity corresponding to the velocity of the cluster members
($\gamma=-6$ km/s, see Fig.~\ref{fig:vrad}) give values for the mass function
$f(m)=(a_{2} \sin(i))^{3}/P^{2}=(m_{1} \sin(i))^{3}/(m_{1}+m_{2})^{2}$ which
are far too large: the values for $m_{1}$ (B-type component) are larger than
60 $\mathrm{M_{\odot}}$, if we assume an $m_{2}$ of about 25 $\mathrm{M_{\odot}}$.\\
If the data obtained in 1908 and 1911 are excluded (these are 20.3 \AA~/mm spectra taken
on photographic plates), the remaining velocities are compatible with a constant
velocity (taking into account the pulsation), or with a small amplitude orbit. But
the $\gamma$ velocity would be close to -16 km/s instead of -6 km/s (cluster velocity).
Furthermore, it is noteworthy that our recent data have a mean value ($-16.7 \pm 2.6$ km/s)
in good agreement with the mean value of the data obtained in 1946 and 1947 
($-15.8 \pm 4.2$ km/s), suggesting that the systematic velocity of the system is
indeed 10 km/s lower than the mean velocity of the cluster.\\

In conclusion, our radial velocity analysis allows to exclude the membership of
\object{V810 Cen} to the cluster \object{Stock~14}.


\section{The blue companion}

The existence of a blue companion was suggested by Parsons \& Peytremann (\cite{parsons1})
and van Genderen (\cite{vangenderen1}) from photometric data and conclusively proved by
Eichendorf et al. (\cite{eichendorf2}) and Parsons (\cite{parsons2}) on the basis of IUE spectra.

From Si IV and C IV line profiles and intensity analysis, these
authors conclude that the spectral type of this blue companion is B0-B1
Iab-Ib, i.e. a star of absolute magnitude $M_{\mathrm{V}}\simeq -6$. Furthermore, 
Parsons (\cite{parsons2}) fitted synthetic spectrum on IUE data and UBVRIJKL photometry and
determined a V magnitude difference between the two components of
$\Delta V \simeq 3.2 $ (the blue component being fainter).
As noted by Parsons, there is an inconsistency between \object{V810~Cen} membership to 
the cluster \object{Stock~14} and the luminosity class of the companion.
Indeed, the absolute magnitude of the two components would be -7.8 
(\object{V810~Cen}) and -4.6 (blue companion) if the membership is 
accepted and the value $\Delta V=3.2$ taken into account. 
In that case, the luminosity of the blue companion corresponds to a giant 
and not to a supergiant (as it is indicated by Si IV and C IV lines). 
In order to reconcile spectral and cluster luminosity, Turner (\cite{turner}) 
suggested that the B star stellar wind could be abnormally strong for
its luminosity, as it is the case for \object{$\tau$ Sco} (B0\,V) and \object{$\xi$ Oph}
(O9 V). If the same phenomenon applies here, this star could be a
giant, with $M_{\mathrm{V}}\simeq -4.6$, in agreement with the membership to 
\object{Stock~14}.

However, as noted in the previous section, the radial velocity analysis do not support
the membership to \object{Stock~14}. Thus, it appears to us that the most logical solution
is to accept~: {\it i)} the luminosity class of the blue companion, probably a
giant B0 star (see section 4), and its absolute magnitude $M_{\mathrm{V}}\simeq -5.1$ 
(Schmidt-Kaler, \cite{schmidt-kaler}); 
{\it ii)} the $V$ magnitude difference between the two components, $\Delta V \simeq 3.3$ 
(see next section). With these values, the absolute magnitude of \object{V810~Cen} is
$M_{\mathrm{V}}\simeq -8.4$ and the star is located behind the cluster \object{Stock~14}.


\begin{figure}
  \resizebox{\hsize}{!}{\includegraphics{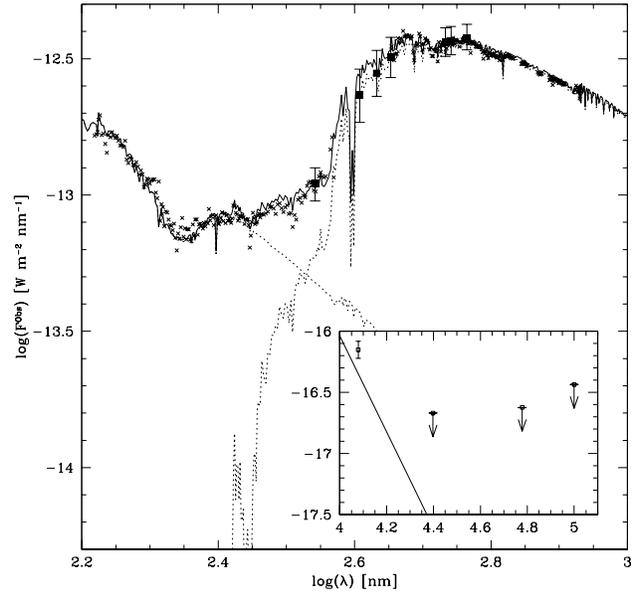}}
  \caption{Reddened and scaled Kurucz models (dashed lines) assuming a B0\,III
    hot companion. The full line is the sum of the B0\,III and F8\,Ia Kurucz model. 
    IUE and Kiehling's data (crosses), the flux derived from Geneva
    photometry (filled squares) and IRAS flux (open squares in the insert)
    are also given. The error bars are the upper and lower flux
    observed in Geneva photometry whereas for IRAS data they represent the
    estimated error. The arrows indicate upper limits for the flux.}
  \label{fig:flux_dist}
\end{figure}


\begin{figure}
  \resizebox{\hsize}{!}{\includegraphics{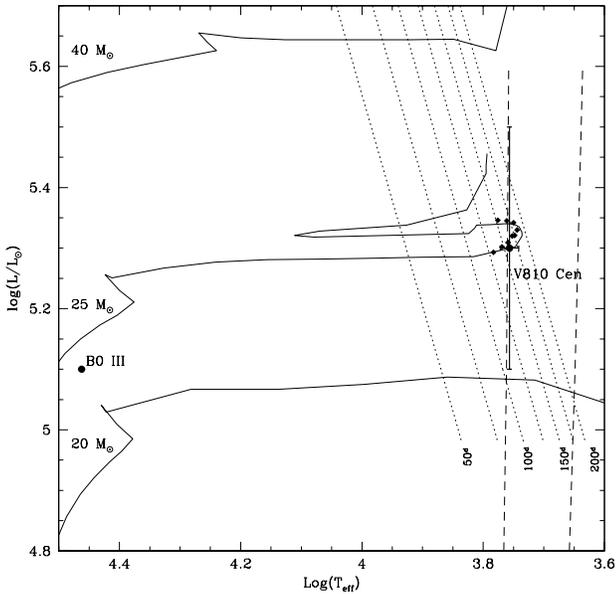}}
  \caption{V810~Cen position in the HR diagram  assuming a B0\,III
    hot companion. The ``error bars'' corresponds to 250 \degr K and 0.5 mag.
    Evolutionary tracks are from Meynet et al. (\cite{meynet}) and the instability strip
    (dashed lines) is from Chiosi et  al. (\cite{chiosi}). Linear non-adiabatic stability
    analysis has been performed on nine models (diamonds) on the 25 $\mathrm{M_{\odot}}$
    track (see also Table ~\ref{tab:lna}). Dotted lines are the iso-periods in the 50 
    to 200 days range (25 days step) for the fundamental mode calculated from
    Schaller (\cite{schaller1}).}
\label{fig:hr_diag2}
\end{figure}


\begin{table*}[hbt]
\begin{tabular}{cccccc}  
\hline \rule{0pt}{2.3ex}
Star	     & Sp. type & $E(B-V)$& $T_{\rm eff}$ & $T_{\rm eff}$ &$\Delta \log T_{\rm eff}$\\
             &  	&       & Evans \& Teays & This paper    &                        \\[0.3ex]
\hline \rule{0pt}{2.3ex}
$\alpha$ Per & F5Iab 	& 0.04	& 6\,270          & 6\,430          & 0.011\\
$\beta$  Aqr & G0Ib  	& 0.03	& 5\,560          & 5\,660          & 0.008\\
9 Peg        & G5Ib	& 0.13  & 5\,110          & 5\,170          & 0.005\\  
\hline 
\end{tabular}

\caption{Temperature estimates for 3 standard supergiants compared to the values 
derived by Evans \& Teays (\cite{evans}). Our temperatures are, in average, 100 K
higher than Evans \& Teays estimates.} 
\label{tab:errors}
\end{table*}

\begin{table*}[hbt]
\begin{tabular}{lccccccc}  
\hline	\rule{0pt}{2.3ex}
Sp	&$T_{\rm eff,1}$ f	&$\log g_{1}$ f	&$C_{1}$	&$T_{\rm eff,2}$	&$\log g_{2}$	&$C_{2}$ &$\chi^{2}$\\[0.3ex]
\hline	\rule{0pt}{2.3ex}
O9 V	&33\,000 		&4.0		&20.286	&6\,070		&1.2		&17.204	&2983\\
B0 V	&30\,000 		&4.0		&20.160	&6\,040		&1.0		&17.198	&3087\\
B1 V	&25\,400 		&4.0		&19.893	&5\,960		&0.4		&17.182	&3654\\
O9 III	&32\,000 		&4.0		&20.249	&6\,060		&1.1		&17.202	&2975\\
B0 III	&29\,000 		&3.5		&20.119	&6\,010		&0.7		&17.190	&2539\\
B1 III	&24\,000 		&3.5		&19.812	&5\,905		&0.0		&17.171	&4236\\
O9 Iab	&32\,600 		&4.0		&20.271	&6\,060		&1.1		&17.203	&2979\\
B0 Iab	&25\,000 		&3.0		&19.898	&5\,910		&0.0		&17.171	&4015\\
B1 Iab	&20\,800	        	&3.0		&19.587	&5\,810		&0.0		&17.151	&9278\\
B2 Iab	&18\,500	        	&3.0		&19.387	&5\,760		&0.0		&17.143	&12405\\
\hline												 
\end{tabular}
\caption{Parameters for various fits to IUE and visible spectra. The microturbulence is set to 4 km/s
and solar metallicity is assumed. The color excess is fixed to $E(B-V)=0.26$. Fixed parameters are quoted 
with an f. The best fit is achieved for a B0 III and F8 Ia couple.} 
\label{tab:models_solar}
\end{table*}

\section{Physical parameters of the two components}
Theoretical energy distributions for each of the two components, taken from the 
grid of stellar atmosphere models of Kurucz (\cite{kurucz}), have been fitted to the 
observed UV, optical and IR spectrophotometric data of \object{V810~Cen} (both components 
together).

In the UV domain, three low dispersion, large aperture, spectra obtained with IUE satellite 
and reduced with NEWSIPS (Nichols \& Linsky, \cite{nichols2}) are available in the IUE Final Archive 
(obtained from NASA Data Archive and Distribution Service), namely lwp6460, lwp28437 and swp26455.
The two lwp spectra have been 
averaged in the 2122-3348 \AA \  range and the strong chromospheric Si IV and C IV lines in 
the swp spectrum have not been used since they cannot be well described by the standard 
Kurucz model atmosphere.

In the optical domain, a 10 \AA \ resolution spectrum is available in the range 3250 to
8600 \AA \ (Kiehling, \cite{kiehling}). These data are accessible through 
the Strasbourg Stellar Data Center. Points flaged as uncertain by the author have not been 
used. In the IR domain, the photometric data from IRAS satellite (IRAS Points Source Catalog, 
\cite{iras}) are reliable only at 12 $\mu m$, whereas the fluxes at 25, 60 and 100 $\mu m$ are only upper limits.
In addition, the photometric measurements in the 7 colors Geneva system
have been used to estimate the effect of the variability. 

No detailed study of the interstellar extinction in the UV domain is available 
towards \object{V810~Cen}. Assuming that the star is not too far away from the open cluster 
\object{Stock~14}, which is at a distance of $2.7 \pm 0.2$ kpc according to Peterson \& FitzGerald (\cite{peterson}),
the mean Galactic extinction law of Kre\l owski \& Papaj (\cite{krelowski}) was adopted. This law is suited 
for stars at 2 to 6 kpc from the sun. 

For each absolutely calibrated flux of the extracted IUE spectra an internal error estimate 
is given in the MXLO files (Nichols et al., \cite{nichols1}). The spectra have been corrected for 
the systematic deviations described by Bohlin (\cite{bohlin2}). From his Fig.~1, the following values 
for the IUE external error have been adopted: 3 \%, 5 \%, 2 \%, and 7 \% respectively in the 
ranges 1100-1900 \AA , 1900-2400 \AA , 2400-3100 \AA \ and 3100-3300 \AA \ . For the optical spectrum, 
Kiehling's external and internal error estimates have been adopted.

To reproduce the observed composite spectrum, with two model atmospheres and the mean 
interstellar extinction law, seven parameters should be estimated: the effective temperature, 
gravity and angular diameter for each star and the color excess. The estimate has been performed 
through a standard least squares procedure minimizing the following quantity:

\begin{equation}
\chi^{2} = \sum_{i} \frac{[\log(F_{i}^{0})-\log(F_{i}^{th})]^{2}}{\sigma_{i}^{2}}
\end{equation}

\noindent
where $F_{i}^{0}$ and $F_{i}^{th}$ are the unreddened observed and theoretical fluxes at the wavelength 
$\lambda_{i}$, and $\sigma_{i}$ is the error estimates on $\log(F_{i}^{obs})$, where $F_{i}^{obs}$ is the 
observed, reddened, flux. To compare the observed and theoretical fluxes the former has been resampled 
at the model frequencies. The theoretical flux is given by:

\begin{eqnarray*}
F_{i}^{th}&=&F_{i}^{mod}(T_{1},g_{1})\;10^{-C_{1}}+F_{i}^{mod}(T_{2},g_{2})\;10^{-C_{2}}\\
\end{eqnarray*}

\noindent
where $F_{i}^{mod}(T,g)$ is the flux from Kurucz's atmosphere models at $\lambda_{i}$ for given values
of $T_{\rm eff}$ and $\log g$. The indices 1 and 2 refer respectively to the blue and red (\object{V810~Cen})
components. The Kurucz's models used are those with the solar abundance and a microturbulence 
of 4 km/s, well suited for supergiants. The parameter $C$ is related to the stellar angular diameter 
$\theta$ through:

\begin{equation}
\theta = 2 \cdot 2.06\:10^{8}\:10^{-0.5\:C}\: \mbox{milliarcsec}    
\end{equation}

\noindent
The unreddened observed flux is given by:

\begin{equation}
\log(F_{i}^{0})=\log(F_{i}^{obs})+\;0.4\;A_{i}
\end{equation}

\noindent
where $A_{i}$ is the interstellar extinction in magnitude, calculated from the extinction law of 
Kre\l owski \& Papaj (\cite{krelowski}), assuming $R=A_{V}/E(B-V)=3.1$.

To estimate the accuracy of the present method, we used the spectrophotometric data
from Glushneva et al. (\cite{glushneva}) for 3 supergiant stars: $\alpha$~Per, $\beta$~Aqr
and 9~Peg. Their parameters have been 
derived by Evans \& Teays (\cite{evans}) on the basis of ultraviolet measurements (IUE data),
optical and infrared photometry (BVRIJHK bands). We have adopted the same values for the 
gravity and the microturbulence as Evans et al. ($\log g$ = 1.5 and $\xi$ =4 km/s 
for the 3 stars). Table ~\ref{tab:errors} shows that our temperatures are slightly larger than
Evans et al. values by 100 K on the average. However, the mean difference is small,
less than 0.01 in $\log T_{\rm eff}$, i.e. less than one spectral subclass. We adopt that value of
100 K for the typical uncertainty of our temperature determinations of \object{V810~Cen}.

According to Eichendorf \& Reipurth (\cite{eichendorf1}) and Turner (\cite{turner}) the blue component 
of \object{V810~Cen} is in the ranges B0-B1 in spectral type and V-Iab in luminosity class. The parameters
$T_{\rm eff,1}$ and $\log g_{1}$ of the blue component were fixed, and the minimization of the 
$\chi^{2}$ value was performed with the free parameters $C_{1}$, $T_{\rm eff,2}$, $\log g_{2}$ and 
$C_{2}$. The computed ranges in spectral type and luminosity class for the blue component were 
O9-B2 and V-Iab. In a first step, the color excess was fixed to 0.26, the mean value for the 
cluster \object{Stock~14}. Table ~\ref{tab:models_solar} shows that the best fit is achieved for a blue 
component of type B0\,III. The corresponding values for the red component are $T_{\rm eff,2} = 6\,010 \mathrm{K}$ 
and $\log g_{2} = 0.7$, corresponding to an F8 supergiant.

In a second step, the color excess was varied from 0.20 to 0.30, for the same parameters of the blue 
component, i.e. those of a B0\,III star. As shown by the $\chi^2$ values in Table ~\ref{tab:E(B-V)_variation}, 
the fitting procedure is weakly dependent on the color excess value~: any value of $E(B-V)$ between
about 0.20 and 0.28 can be accepted.

From the relation~: 

\begin{equation}
log(R_{\rm 2}/R_{\rm 1}) = 0.5 (C_{1} - C_{2})    
\end{equation}

we derive the radius ratio $R_{\rm 2}/R_{\rm 1} = 31.1$ which clearly indicates that
the luminosity class of the red component is Ia. Thus, the best solution for the
two components of \object{V810~Cen} is~:\\ 

Blue component : B0\,III, $T_{\rm eff} = 29\,000 \pm 1\,000$ K

Red component  : F8\,Ia, $T_{\rm eff} = 5\,970 \pm 100$ K \\

This solution is plotted in Fig.~\ref{fig:flux_dist} where reddened model
fluxes (dashed lines) and their sum (full line) are compared to the fluxes in the
7 Geneva passbands (full boxes, computed according to Rufener \& Nicolet, \cite{rufener3}).
Furthermore, IRAS photometry is available for the point source 11410-6212 associated
to \object{V810~Cen} and it is compared to the model in the insert. The error bars, associated
with Geneva photometry, are the maximum and minimum observed fluxes, while those plotted
at 12 $\mu \rm m$ are the 16 \% uncertainties quoted in the IRAS point source
catalogue. Those data marked by an arrow are upper limits fluxes.

The absolute and bolometric magnitudes have been derived from the spectral type (Lang, \cite{lang})
for the blue component and using $\Delta V =3.3$ for the red component (\object{V810~Cen}). The values 
of the stellar radii have been calculated from $T_{\rm eff}$ and $M_{\rm bol}$. Thus we have~:\\

Blue comp.~: $M_{\rm V}=-5.1, M_{\rm bol}=-8.0, R=14$ R$_{\odot}$

Red comp.~: $M_{\rm V}=-8.4, M_{\rm bol}=-8.5, R=420$ R$_{\odot}$ \\

Note that the ratio $R_{\rm 2}/R_{\rm 1}$ is 29.8, very close to the value derived 
above from the values of the parameter C. The two components can then be placed 
in the $\log L$ vs. $\log T_{\rm eff}$ diagram (see Fig.~\ref{fig:hr_diag2}) together with
the evolutionary tracks from Meynet et al. (\cite{meynet}). Initial masses of $25 \pm 5$ M$_{\odot}$ can 
be derived for both components.

As already noted by Eichendorf \& Reipurth (\cite{eichendorf1}), \object{V810~Cen} is located near the blue edge
of the instability strip (in Fig.~\ref{fig:hr_diag2} the limits of the strip for the fundamental
mode, dashed lines, are from Chiosi et al. (\cite{chiosi})).

With the values of the absolute magnitude of both components and of the mean visual
apparent magnitude ($V$ = 5.021), we derive for \object{V810~Cen} a distance
of 3.3 to 3.5 kpc (for E(B-V) in the range 0.28 to 0.24). Thus \object{V810~Cen} is located
0.6 to 0.8 kpc behind the stellar cluster \object{Stock~14}. Is this location 
in contradiction with the color excess values (0.26 for \object{Stock~14}, 0.20-0.28 for \object{V810~Cen})~?
The answer is no, because \object{Stock~14} is non-uniformly reddened. For the member stars of the cluster,
Peterson \& FitzGerald (\cite{peterson}) found a standard deviation of 0.021 on the individual values of
$E(B-V)$. This value corresponds to a $3 \sigma$ interval in $E(B-V)$ of 0.19-0.32. Moreover,
the two stars of \object{Stock~14} which have the lowest reddening, $E(B-V)$=0.18 (star 26) and
0.21 (star 27) according to Turner (\cite{turner}), are located in the sky very close to \object{V810~Cen}
(stars 33, 34 and 35, are also close, but cannot be used because of their unreliable photometric data).


\begin{table}[hbt]
\begin{tabular}{ccccccccc}  
\hline	\rule{0pt}{2.3ex}
E(B-V)	&$\chi^{2}$	&$C_{1}$	&$T_{eff,2}$	&$\log(g_{2})$	&$C_{2}$	\\[0.3ex]
\hline \rule{0pt}{2.3ex}
0.20	&2795	&20.314	&5\,890	&1.1	&17.216\\
0.22	&2581	&20.249	&5\,930	&1.0	&17.207\\
0.24	&2495	&20.184	&5\,970	&0.9	&17.199\\
0.26	&2639	&20.119	&6\,010	&0.7	&17.190\\
0.28	&2721	&20.054	&6\,040	&0.6	&17.181\\
0.30	&3069	&19.988	&6\,070	&0.5	&17.171\\
\hline												 
\end{tabular}

\caption{Variation of the $\chi^{2}$ with respect to $E(B-V)$ if a B0\,III 
companion is assumed (see Table ~\ref{tab:models_solar}). The microturbulence is set to 4 km/s 
and solar metallicity is assumed.} 
\label{tab:E(B-V)_variation}
\end{table}


\begin{figure}
\resizebox{\hsize}{!}{\includegraphics{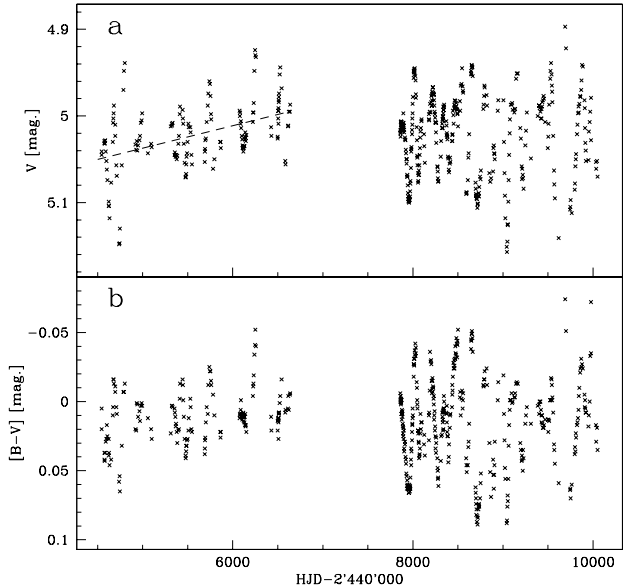}}
\caption{Geneva photometric data for V and [B-V]. The dashed line in figure \emph{a}
shows the long-term brightening of $0.0091$ mag/y.}
\label{fig:light_curves}
\end{figure}

\section{The photometric data}

\object{V810~Cen} was measured 512 times in the seven filters of the Geneva Photometric system (Golay, \cite{golay};
Rufener, \cite{rufener2}) from JD 2\,444\,544 (Nov. 1980) to 2\,450\,051 (Dec. 1995) with a lack of observations 
from 2\,446\,638 to 2\,447\,861 (see Fig.~\ref{fig:light_curves}). 
These data come from the successive Swiss telescopes (40~cm and 70~cm) at ESO La~Silla Observatory (Chile),
equipped with the P7 photoelectric photometer (Burnet \& Rufener, \cite{burnet}). The data reduction has been
made according to the method described by Rufener (\cite{rufener3}). Only the 499 measurements having weights in
magnitude (Q) and in colors (P) larger than 0 have been used. 

The photometric data are available in electronic format at the CDS via anonymous
ftp to cdsarc.u.strasbg.fr (130.79.128.5) or via http://cdsweb.u-strasbg.fr/Abstract.html

Table ~\ref{tab:mean_measure} displays the mean magnitude values and rms for the 7 filters 
for \object{V810~Cen} and 3 standard stars. The corrected rms value $\sigma_{0}$ for \object{V810~Cen} is given by
$\sigma_{0}^2 = \sigma^2 - [(1/3)(\sigma_{s1} + \sigma_{s2} + \sigma_{s3})]^2$, where $\sigma$ is the
rms of \object{V810~Cen} ($\sigma_{0}$ differ from $\sigma$ by less than 0.001). At this stage, the essential 
characteristics of the variability of \object{V810~Cen} is~: {\it i)} an increase of $\sigma_{0}$, thus of the 
global amplitude from G to B1 filters; {\it ii)} an amplitude in U smaller than in B1, B and B2. This 
apparent decrease of the amplitude in U is due to the constant flux from the B-type component, which
contribute significantly to the total flux only in this filter.

\begin{table*}[hbt]

\caption{Weighted mean value and rms of the 499 measurements of the 7 magnitudes of \object{V810~Cen} in the Geneva 
Photometric System. The rms is also given for 3 standard stars measured during the same nights than 
V810~Cen. }
\label{tab:mean_measure}

\begin{tabular}{lcccccccc}  
\hline 
           &Filter   		        	&$U$	&$B1$   &$B$	&$B2$	&$V1$   &$V$	&$G$\\
           &$\lambda_{0}\: [$\AA$]$   	        &3464	&4015   &4227	&4476	&5395   &5488	&5807\\[0.3ex]
\hline 
V810~Cen   &Mean mag.    		        &6.797	&6.155  &5.035	&6.323	&5.785  &5.021	&6.005 \\
V810~Cen   &corrected rms ($\sigma_{0}$)        &0.055	&0.081 &0.070	&0.062	&0.046 &0.044	&0.040\\[0.3ex]
\hline 
HD~93540   &rms ($\sigma_{s1}$)			&0.0070	&0.0061 &0.0057	&0.0058	&0.0050 &0.0043	&0.0048\\
HD~94510   &rms ($\sigma_{s2}$)		      	&0.0083	&0.0065 &0.0060	&0.0058	&0.0048 &0.0040	&0.0048\\
HD~102350  &rms ($\sigma_{s3}$)		       	&0.0068	&0.0056 &0.0047	&0.0049	&0.0044 &0.0035	&0.0046\\
\hline				                                	
\end{tabular}


\end{table*}

Our monitoring of \object{V810~Cen} can be characterized as follows~:

-- Three measurements have been obtained in February and March 1976, and in February 1977 (HJD
from 2\,442\,817 to 2\,443\,182).

-- From October 1980 to July 1986 (HJD from 2\,444\,544 to  2\,446\,639), the air mass of the 
measurements was restricted to values smaller than about 1.7. Thus the star was measured
each year, only between November and July. On the average, 25 measurements per year were
obtained.

-- From December 1989 to October 1991 (HJD from 2\,447\,860 to  2\,448\,559), a peculiar observational
effort was done in order to have a monitoring as continuous as possible (the star
is observable during the whole year from La~Silla). On the average, 9 measurements per month
were obtained. These data are shown in Fig.~\ref{fig:wwz}a. Variations with a characteristic time in the 
100-150 days range are clearly exhibited. In addition, the amplitude is variable.

-- From November 1991 to December 1996 (HJD from 2\,448\,590 to  2\,450\,051), the monitoring was less 
intensive. Due to several periods without data, the average number of measurements per month is 3.

\section{The photometric variability}

Three methods have been used to analyze the variability of \object{V810~Cen}~:\\

-- The Date Compensated Discrete Fourier Transform {\bf DCDFT} described by 
Ferraz-Mello (\cite{ferraz-mello}). This method avoids the missmeasurement of the amplitudes
and mean magnitude encountered when the classical Discrete Fourier Transform
(Deeming, \cite{deeming}) is applied to unevenly spaced data (see Foster,
\cite{foster1}, \cite{foster2}).\\

-- The {\bf CLEANEST} method of Foster (\cite{foster1}). The spurious peaks in the Fourier analysis induced by the
data sampling are eliminated by an iterative deconvolution process. The iteration has been stopped when the 
residual peaks in DCDFT were below the 1\% confidence level limit according to Foster (\cite{foster2}).\\

-- The Weighted Wavelet Z-transform method ({\bf WWZ}) of Foster (\cite{foster3}).
Schematically, the data are weightened with a gaussian
function (with $\sigma\simeq$ 200 d to 300 d in our present case) centered at a given time $\tau$.
The Fourier Transform is then performed for successive $\tau$ values and the WWZ maximum 
($\mathrm{WWZ}_{\mathrm{max}}$) indicates the dominating frequency $\nu(\mathrm{WWZ}_{\mathrm{max}})$
(noted hereafter $\nu_{\mathrm{max}}$).
The amplitude of $\nu_\mathrm{max}$ is estimated by the means of 
the Weighted Wavelet Amplitude statistics ($\mathrm{WWA}_{\mathrm{max}}$).
The WWZ method allows to analyze the changes in amplitude or/and frequency of the light curve.\\

In order to check the period and amplitude variabilities, the data have been splited into two samples: sets 1
and 2 span the ranges in HJD 2\,444\,544--2\,446\,639 and  2\,447\,860--2\,450\,051 respectively 
(see Fig.~\ref{fig:light_curves}).
A careful examination of the data shows that the observed variations cannot be simply described
as a stable, multiperiodic function. The main characteristics of the variability of
\object{V810~Cen} during our survey are the following~:\\

-- {\bf Long-term variation}~: the mean luminosity of the star increased regularly from 
February 1976 to July 1986 (see Fig.~\ref{fig:light_curves}). The rate of brightening was roughly 
$0.009 \:$ mag/year. Since November 1991, the mean luminosity remained constant.\\

\begin{figure}
\resizebox{\hsize}{!}{\includegraphics{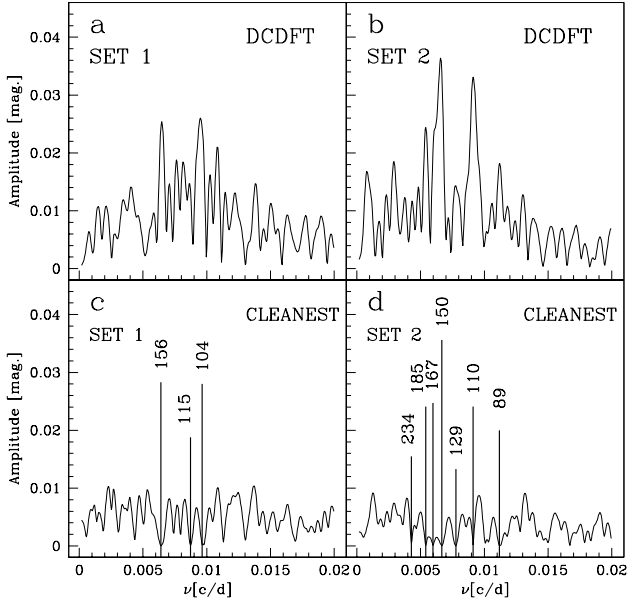}}
\caption{\emph{a and b}: Date-compensated discrete Fourier transform (DCDFT) amplitudes for
sets 1 and 2. \emph{c and d}: CLEANEST for sets 1 and 2, it is a combination of the subtracted
periods and amplitudes (vertical bars each labeled with their period) and the DCDFT of the residuals (full line).
CLEANEST has been applied until all the peaks of the DCDFT \emph{power} are below the  1 \% confidence level 
limits (computed according to Foster, \cite{foster2}).}
\label{fig:cleanest}
\end{figure}

-- {\bf Modes at $\sim$156 and $\sim$107 days}~: Figs. ~\ref{fig:cleanest}a and b present
the DCDFT for both sets. We note two main peaks at frequencies 0.0064 and 0.0093 d$^{-1}$.
The frequency difference is close the 1~y$^{-1}$ alias ($\simeq 0.00274$~d$^{-1}$), suggesting
that one of these may be an alias. However both of them are real because they are kept
by the CLEANEST iterative method.\\

-- {\bf Other modes}~: the CLEANEST spectra (Figs. ~\ref{fig:cleanest}c and d) reveal a third period at 115 d for
set 1, and five new periods at 89, 129, 167, 185 and 234 days for set 2. These peaks appear thanks to the better
time sampling of the second set. \\

A curve of the form~:

\begin{equation}
f(t) = \sum_{i=1}^{n} A_{i} \, \cos [2\pi \nu_{i}(t-t_{0}) +\phi_{i}]
\end{equation}

\noindent has been fitted to the observations in sets 1 and 2 and the resulting amplitudes are given
in Table ~\ref{tab:cleanest}. The residual standard deviation is 0.019 mag for both sets,
a quite large value compared to the accuracy of our measurements (0.004 mag, see Table 
~\ref{tab:mean_measure}). As we shall see in the following, this is due to amplitude 
or/and period variation.\\


\begin{table}[hbt]
\caption{Frequency, period and amplitude  estimate from CLEANEST for
set 1 (HJD 2\,444\,544--2\,446\,639) and 2 (HJD 2\,447\,860--2\,450\,051).}
\label{tab:cleanest}

\begin{tabular}{cccc}  
\hline	\rule{0pt}{2.3ex}
&Frequency [c/d]    &Period [d]	 &Amplitude [mag.]  \\[0.3ex]
\hline \rule{0pt}{2.3ex}
Set 1	&0.00641$\pm$2e-05&156.0$\pm$0.5&0.028$\pm$0.002\\
	&0.00964$\pm$2e-05&103.7$\pm$0.2&0.028$\pm$0.002\\
	&0.00873$\pm$3e-05&114.6$\pm$0.4&0.019$\pm$0.002\\[0.3ex]
\hline \rule{0pt}{2.3ex}
Set 2	&0.00668$\pm$2e-05&149.8$\pm$0.2&0.036$\pm$0.002\\
	&0.00912$\pm$3e-05&109.6$\pm$0.2&0.024$\pm$0.002\\
	&0.01119$\pm$3e-05& 89.4$\pm$0.2&0.020$\pm$0.002\\
	&0.00541$\pm$3e-05&184.8$\pm$0.5&0.024$\pm$0.002\\
	&0.00597$\pm$2e-05&167.4$\pm$0.4&0.025$\pm$0.002\\
	&0.00428$\pm$4e-05&233.5$\pm$1.3&0.015$\pm$0.002\\
	&0.00777$\pm$5e-05&128.7$\pm$0.5&0.013$\pm$0.002\\
\hline 
\end{tabular}

\end{table}


\section {Analysis of the mode variations}

\begin{figure}
\resizebox{\hsize}{!}{\includegraphics{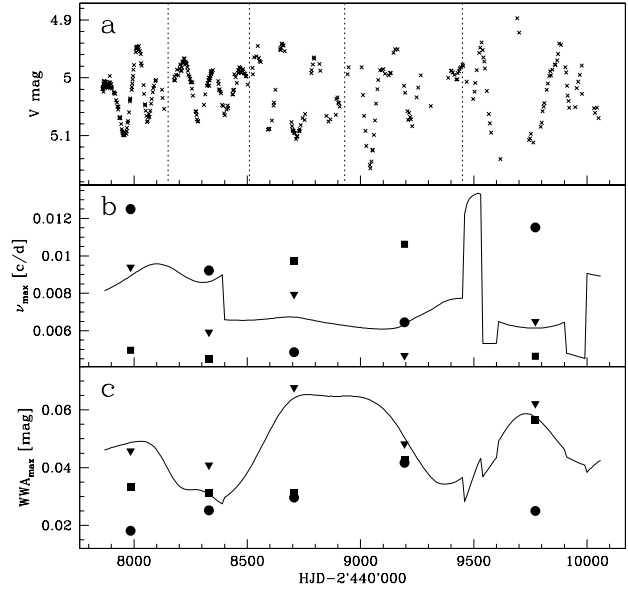}}
\caption{\emph{a}: V-band photometry for data set 2, with subset division (vertical dotted lines). 
\emph{b and c}: dominating frequencies (WWZ maximum) for a given ``mean time'' $\tau$ and its
corresponding amplitude WWA (full lines). 
Triangle, square and dot symbols (sorted according to decreasing amplitude) are the tree main 
frequencies found with DCFT analysis for each subset shown in \emph{a}.}
\label{fig:wwz}
\end{figure}

The data density is large enough in set 2 (see Fig.~\ref{fig:wwz}a) to allow an
analysis of the variations of the frequencies and/or amplitudes of the dominant 
modes. Two kinds of analysis have been done~:\\

-- DCDFT method is performed in five successive intervals. In each of them 
(see Fig.~\ref{fig:wwz}a), the
three most important modes have been identified and their frequencies and amplitudes are
plotted as full symbols in Figs. ~\ref{fig:wwz}b and c (the decreasing amplitude are
symbolized with triangles, squares and dots respectively).

-- WWZ method described previously is applied to the entire interval. In our case, the
WWZ method gives reliable results only for the mode with the largest power 
($\mathrm{WWZ}_{\mathrm{max}}$). The variations of the dominating frequency
$\nu_{\mathrm{max}}$ and its amplitude $\mathrm{WWA}_{\mathrm{max}}$ are shown
as solid lines in Figs.~\ref{fig:wwz}b and c.\\

The main points revealed by this analysis are~:

\begin{enumerate}

\item The mode at $\sim$107 d ($\sim 0.0093$ d$^{-1}$) is the most important one before 
HJD 2\,448\,400, while the mode at $\sim$156 d ($\sim 0.0064$ d$^{-1}$) dominates afterwards
(with short exceptions around HJD 2\,449\,500, 2\,449\,580 and 2\,449\,950).
At the end of our survey, it appeared that the mode at $\sim$107 d was again the most 
significant one (see Fig.~\ref{fig:wwz}b).

\item The amplitude of the most important mode varies almost continuously. 
However, a stable amplitude at $\sim$0.06 mag was observed in the HJD interval 
2\,448\,700--2\,449\,100, for the mode at $\sim$156 d (see Fig.~\ref{fig:wwz}c).

\item The DCDFT analysis (Fig.~\ref{fig:wwz}b) shows that a mode at $\sim 0.0045-0.0055$~d$^{-1}$
(period between 225 and 180~d) is always present, but is never the most important one.

\end{enumerate}

In conclusion, {\bf \object{V810~Cen} is a multi-periodic small amplitude variable star, 
whose periods and amplitudes are variable with time}. The periods of the two dominant
modes are $\sim$156 and $\sim$107 d. The 7 years intensive survey (1989-1996) also
reveals a third mode, with a period in the range 180-225 d.


\section {The light and color curve parameters}

The amplitude ratios and the differences in phase between the light $V$ and color $[B-V]$ curves are interesting
for the variability description, in particular for the mode identification. Because of the multiperiodic
character of the light curve and of the amplitude variation described in the previous sections, the
amplitude and phase of the principal modes have been fitted in various time intervals through equation (5),
setting the number of frequencies (n) to 3.

\begin{figure}
\resizebox{\hsize}{!}{\includegraphics{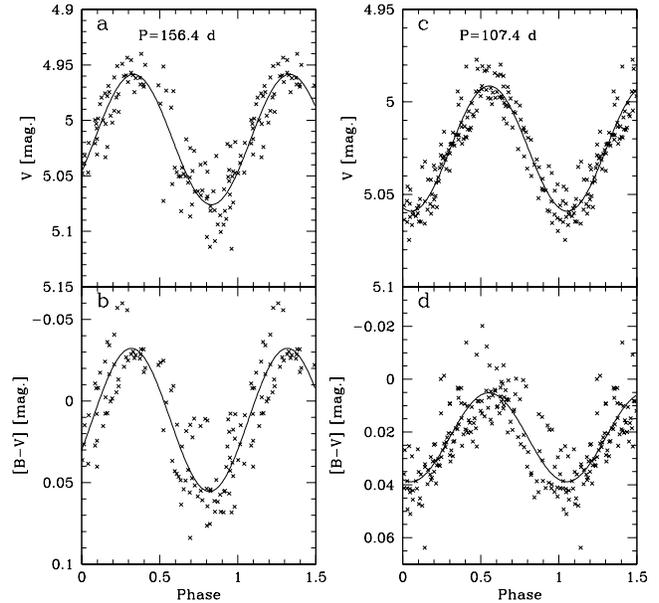}}
\caption{Light and color curves of the two main modes at 107.4~d (time interval: HJD 2\,447\,850-2\,448\,400)
and 156.4~d (time interval: HJD 2\,448\,400-2\,449\,400).}
\label{fig:v_bv_modes}
\end{figure}

According to Fig.~\ref{fig:wwz}, the parameters of the mode at $\sim$107, $\sim$156 and $\sim$187 d have
been calculated respectively in the time intervals in HJD 2\,447\,850--2\,448\,400, 2\,448\,400--2\,449\,400 and 
2\,447\,850--2\,450\,100.
The results are given in Table ~\ref{tab:magcol}. The light and color curves for the two principal modes at
$\sim$107 and $\sim$156 days are shown in Fig.~\ref{fig:v_bv_modes}. As we see, the light and color variations are clearly
{\it in phase}, i.e. the star is bluer when brighter. This is an indication of cepheid-like pulsation nature.
However, it must be noted that, taken into account the uncertainties on $\Delta\phi$ in Table ~\ref{tab:magcol},
the values of the phase difference $\phi_{V} - \phi_{B-V}$ can be slightly negative, zero or slightly positive.
According to Balona \& Stobie (\cite{balona2}, \cite{balona3}), this indicates that these two modes can be radial
($\Delta\phi < 0$), but could also be non-radial, with odd ($\Delta\phi = 0$) or even ($\Delta\phi > 0$) 
values of the spherical harmonic order $l$.

The third mode, at $\sim$187 days, shows a clearly positive value of $\Delta \phi$. This might indicate a non-radial
quadrupole mode ($l=2$). However, due to the amplitude and period changes in the modes of \object{V810~Cen}, the
mode identification must not be made on the basis of $\Delta\phi$ values only (see next section).

\begin{figure}
\resizebox{\hsize}{!}{\includegraphics{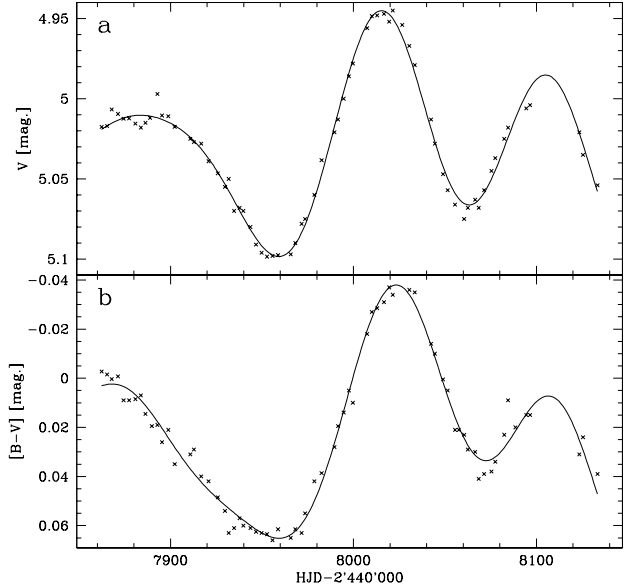}}
\caption{Three-frequency fit (0.00942 c/d, 0.00488 c/d and 0.01232 c/d) 
for $V$ and $[B-V]$ (\emph{a} and \emph{b} respectively). Data are resampled in 
3 days wide bin to homogenize the sampling.}
\label{fig:v_bv}
\end{figure}

When the time interval is short enough and the monitoring sufficiently dense, very good fitted light and color 
curves can be achieved. This is illustrated in Fig.~\ref{fig:v_bv}, for the interval in HJD 2\,447\,850--2\,448\,150 
(the first interval in Fig.~\ref{fig:wwz}a). The solid curve in Fig.~\ref{fig:v_bv} represents a fit
with 3 modes, the most important one being at $\sim107$ d. The residual standard deviation of the observed
values around the fitted curves are 0.0050 mag in $V$ and 0.0046 mag in $[B-V]$. Thus, in that case, the three modes 
describe completely the variability of \object{V810~Cen} (see Table~\ref{tab:mean_measure}).


\begin{table*}[hbt]
\caption{Parameters of the highest amplitude variability modes according to equation (5) in various time intervals.}
\label{tab:magcol}

\begin{tabular}{cccccccc}  
\hline \rule{0pt}{2.3ex}
Period [d] & Frequency [c/d] & HJD interval & $V$ Amp. $A_{V}$ & $[B-V]$ Amp. $A_{B-V}$& $A_{V}/A_{B-V}$ & $\Delta\phi=\phi_{V}-\phi_{B-V}$ \\[0.3ex]
\hline \rule{0pt}{2.3ex}
107.4 & 0.00931 & 2\,447\,850-2\,448\,400 & 0.034 $\pm$0.001 & 0.017$\pm$0.001 & 2.01 & -0.003 $\pm$0.014\\
156.4 & 0.00639 & 2\,448\,400-2\,449\,400 & 0.059 $\pm$0.003 & 0.044$\pm$0.003 & 1.34 & -0.014 $\pm$0.018\\
186.7 & 0.00536 & 2\,447\,850-2\,450\,100 & 0.022 $\pm$0.002 & 0.014$\pm$0.002 & 1.52 &  0.071 $\pm$0.038\\
\hline 
\end{tabular}

\end{table*}


\section{The origin of the variability}

\subsection{Comparison with supergiants and cepheids calibrations}

\begin{itemize}

\item {\bf The Period--Luminosity--Colour PLC relations}. The largest amplitude period
($\sim$156~d) is in very good agreement with the empirical
PLC relations for yellow supergiants. Taking the parameters derived in Section~4, 
$M_{\rm bol} = -8.5$ and $T_{\rm eff} = 5970$~K, the relation from Maeder \& Rufener (\cite{maeder}),
$\log P = -0.346 M_{\rm bol} -3 \log T_{\rm eff} + 10.60$,
gives $P = 163$~d. Furthermore, assuming an evolutionary mass of $\sim$20 M$_{\odot}$
(see the end of this Section), the relation from Burki (\cite{burki1}), 
$\log P = -0.38 M_{\rm bol} -3 \log T_{\rm eff} -0.5 \log M + 10.93$,
gives $P = 152$~d.\\
The agreement is not so good with the 
Period--Luminosity relation for galactic cepheid stars (Gieren et al., \cite{gieren2})~: $M_{\rm V} = -1.371 - 2.986 \log P$
(scatter~: 0.26). Using the $M_{\rm V}$ value obtained in Section~4, the predicted period is 226~d. This 
discrepancy with the observed period ($\sim$156~d) is quite normal since \object{V810~Cen} is not located in the
center of the cepheid instability strip, but rather in its blue border. For that reason, a color term
is necessary in the PLC relation for supergiants (see Burki, \cite{burki1}). Thus, {\bf the observed main 
mode is in agreement with the observations concerning pulsating supergiants}.

\item {\bf The Period--Radius relation for cepheids}. Gieren et al. (\cite{gieren1})
have established such a relation
on the basis of 101 classical cepheids in the period range 3--45~d, studied from the visual
brightness technique~: $\log R = 1.108 + 0.743 \log P$ (scatter $\sigma$ = 0.071). It is noteworthy
that, despite the very large value of its main period ($\sim$156~d), which is well outside the
period range covered by the classical Galactic cepheids, the radius of \object{V810~Cen} found in Section~4
($\log R$ = 2.62) is in agreement with this global relation for cepheids ($\log R$ = 2.74), 
the difference being only 1.6~$\sigma$. This agreement is {\bf a strong support for a 
cepheid--like pulsation of this main mode, i.e. a fundamental radial mode}.

\end{itemize}

\subsection{Theoretical periods of the the radial modes} 

\begin{itemize}

\item {\bf Stellar models with ``standard'' mass loss rate}. Lovy et al. (\cite{lovy}) and 
Schaller (\cite{schaller1})
have calculated the pulsation periods for the fundamental radial mode and the first two overtones, from
a linear non-adiabatic analysis. Schaller's analysis is based on the massive supergiant models from the grid
of evolutionary tracks of massive stars by Schaller et al. (\cite{schaller2}). In these
models, the dependence of the mass
loss rate on $L$ and $T_{\rm eff}$ is that given by de Jager et al. (\cite{dejager}),
called hereafter the ``standard'' mass loss rate. Schaller (\cite{schaller1}) derived a
theoretical PLC relation, $\log P = -0.38 M_{\rm bol} -3 \log T_{\rm eff} + 10.60$, 
for the fundamental radial mode which predicts a period $P_0$ = 140~d for \object{V810~Cen}. 
This value is valid for models cooler than $\log T_{\rm eff} = 4.1$ and in the phase 
of helium burning. 

\item {\bf Stellar models with ``high'' mass loss rate}. Meynet et al. (\cite{meynet}) have
calculated a new grid of evolutionary stellar models, by adopting a ``high'' mass loss 
rate, i.e. a rate enhanced by a factor of two with respect to the ``standard'' mass loss rate.
Various comparisons with observations of high-mass stars support the high-mass loss grid
of stellar models (see Maeder \& Meynet, ~\cite{maeder3}). The linear non-adiabatic
pulsation code of Schaller (\cite{schaller3}) has been applied to various 25~$\mathrm{M_{\odot}}$ models, 
at the edge of the first redward evolution, close to \object{V810~Cen} position. The result of these
calculations is presented in Table~7 for the lower (redward evolution) and upper (blueward
evolution) tracks. A fundamental mode period of 157~d is obtained for the blueward
evolution at $\log T_{\rm eff}$ = 3.750 and $\log L/L_{\odot}$ = 5.342, that is 0.004 dex away from
values derived in section 4 for \object{V810~Cen} ($\log T_{\rm eff}$ = 3.78, $\log L/L_{\odot}$ = 5.3).
Due to a strong mass loss at the beginning of the blueward evolution, the 
mass has decreased to 20~$\mathrm{M_{\odot}}$. 

\end{itemize}

\subsection{Interpretation of the main modes}

\begin{itemize}

\item {\bf The observed main period at $\sim$156~d}. This observed period is in good
agreement with the theoretical prediction for models with ``high'' mass loss rate (157.2~d).
Thus, {\bf the main observed period of \object{V810~Cen} ($\sim$156~d) can be explained
by a pulsation in the radial fundamental mode of a supergiant with an initial mass close to
$\sim 25 \,{\rm M}_{\odot}$ }. The evolutionary mass of \object{V810~Cen} must be $\sim 20 
\,{\rm M}_{\odot}$. 

\item {\bf The observed second period at $\sim$107~d}. The predicted periods of the 
first two radial overtones ($P_1$ = 82.0~d, $P_2$ = 56.3~d, see Table~7) are much too 
small to explain this period. The period ratio of the observed two dominant modes 
is 0.69, while the predicted ratios are $P_1/P_0$ = 0.52, $P_2/P_0$ = 0.36, $P_2/P_1$ = 0.69.
This last value could suggest that \object{V810~Cen} pulsates in the first and second radial 
overtones. However, in that case, the fundamental radial period would be $\sim$350~d.
This period would require  a very high luminosity, $M_{\rm bol} \simeq -9.5$, which is
difficult to postulate for a F8\,Ia star. Since, in addition, this very large period is not
observed, we conclude that {\bf the observed second main period cannot be a radial mode
and, thus, a non-radial p-mode is the most natural explanation}. 

\item {\bf The observed third period at $\sim$187~d}. This period is larger than the main 
mode at $\sim$156~d. With the hypothesis that the main mode is the radial fundamental 
pulsation, we have to conclude that {\bf the observed third period cannot either be a 
radial mode and, thus, a non-radial g-mode is to be postulated}. Non-radial g-modes have
already been suggested in supergiants by Maeder (\cite{maeder2}) or de Jager et al. 
(\cite{dejager2}), and are probably present in the case of yellow hypergiants like $\rho$~Cas
(Lobel et al., \cite{lobel}). 

\end{itemize}

As noted above, the period ratio of the observed two dominant modes is 0.69, thus very close 
to the values of the double-mode cepheids pulsating in the radial fundamental mode and first 
overtone with $P_1/P_0$ in the 0.696--0.711 range (Balona, \cite{balona1}). However, \object{V810~Cen}
has a high mass, and therefore its $P_1/P_0$ value is lower than the ratio observed for ``low mass'' 
double-modes cepheids (see table ~\ref{tab:lna}). The similarity between the observed value of 0.69 and
the 0.696--0.711 range for double-modes cepheids is at the origin of a wrong conclusion, 
i.e. \object{V810~Cen} is a double-mode cepheid in addition to its supergiant
characteristics (Burki, \cite{burki2}).

An alternative interpretation associating the $\sim$187~d period to the radial fundamental has also been 
considered; then the other modes would be non-radial p-modes. However, this alternative is unlikely
because, if the fundamental mode is excited, it should have the highest amplitude and this is not 
the case for the $\sim$187~d period (the highest amplitude mode is the $\sim$156~d period). Furthermore,
the (marginally significant) positive phase shift does not support that hypothesis neither and there
are strong evidences for the {\em radial nature} of the $\sim$156~d period.

Our present conclusion, based on a very long-term photometric monitoring, is that \object{V810~Cen} could be
a supergiant star exhibiting the three types of pulsation modes~: {\bf radial fundamental mode} 
(main mode at $\sim$156~d), {\bf non-radial p-mode} (second mode at $\sim$107~d) and {\bf non-radial
g-mode} (third mode at $\sim$187~d). If this conclusion is correct, \object{V810~Cen} would be the 
first known case of a star showing such a diversity of pulsational characteristics.

\begin{table*}[hbt]
\caption{Periods and growth rates ($\eta$) from linear non-adiabatic stability analysis of 
stellar model with ``high'' mass loss rate. Positive $\eta$ indicate unstable modes.}
\begin{tabular}{ccccccccccc}  
\hline \rule{0pt}{2.3ex}
&$\log(T_\mathrm{eff})$&$\log(\frac{L}{L_{\odot}})$&$\frac{M}{\mathrm{M_{\odot}}}$&$P_{0}$ [d]    &$\eta_{0}$	 &$P_{1}$ [d]    &$\eta_{1}$	 &$P_{2}$ [d]    &$\eta_{2}$&$P_{1}/P_{0}$\\[0.3ex]
\hline \rule{0pt}{2.3ex}
Redward		&3.783& 5.293& 22.3& 107.88&-0.213& 50.92& 0.134&  33.29& -0.346& 0.47\\
evolution       &3.769& 5.302& 22.2& 112.51&-0.018& 56.93& 0.141&  39.92& -0.176& 0.51\\
                &3.759& 5.309& 22.2& 123.04& 0.008& 64.54& 0.146&  44.67& -0.159& 0.52\\
                &3.752& 5.320& 22.1& 129.62& 0.040& 69.98& 0.148&  47.77& -0.177& 0.54\\
                &3.748& 5.321& 21.7& 132.03& 0.068& 72.66& 0.143&  49.37& -0.192& 0.55\\[0.3ex]
\hline \rule{0pt}{2.3ex}		    	   
Blueward	&3.744& 5.330& 20.6& 137.43& 0.095& 76.51& 0.130&  52.11& -0.224& 0.56\\
evolution       &3.750& 5.342& 16.6& 157.17& 0.127& 81.99& 0.090&  56.25& -0.489& 0.52\\
                &3.761& 5.345& 14.9& 165.45& 0.246& 79.22& 0.052&  53.16& -0.829& 0.48\\
                &3.776& 5.346& 13.7& 179.70& 0.041& 79.33& 0.230&  52.18& -1.900& 0.44\\
\hline 
\end{tabular}
\label{tab:lna}
\end{table*}

\section{Conclusion}

Radial velocities, spectrophotometry and Geneva photometry data have been used to
discuss the evolutionary status and pulsation modes of the yellow variable supergiant \object{V810~Cen}
and its B-type companion.

Radial velocities have been collected from the \textsc{Coravel} database as well as from the literature.
They do not support the \object{Stock~14} membership hypothesis of \object{V810~Cen}. To be consistent with the 
spectrophotometric data, the star has to be beyond \object{Stock~14}, at 3.3 to 3.5 kpc from the sun.

IUE and visible spectrophotometry together with Kurucz atmosphere models have been used to estimate 
the physical parameters of the red supergiant~: $T_{\rm eff} = 5\,970 \pm 100$ K, 
$M_{\rm V}=-8.4, M_{\rm bol}=-8.5, R=420$ R$_{\odot}$, spectral type F8~Ia. This star is about 
3.3 mag brighter than its B0\,III companion. Thus, the observed variability is due to the red supergiant.

The derived temperature and luminosity place the star at the blue edge of the classical 
cepheids instability strip. The initial mass was $\sim 25\, {\rm M}_{\odot}$ while the evolutionary mass  
ought to be close to $\sim 20\, {\rm M}_{\odot}$.

The high-precision long-term photometric monitoring in the Geneva system reveals various types of 
variability~:

\begin{itemize}
\item A long-term increase of the luminosity from February 1976 to July 1986.

\item A multimode pulsation behavior, with periods and amplitudes variable with time (in the range 0.02 to 0.06 mag). 
The main modes are~:

  \begin{itemize}

  \item A dominant $\sim$156~d period, identified as the fundamental radial mode.
        Its period value is in good agreement with the fundamental radial period derived from 
	theoretical non-adiabatic linear pulsation analysis. 
  
  \item The second mode with a period of $\sim$107~d, which is most probably a non-radial mode, may be a p-mode.

  \item The third mode with a period of $\sim$185~d, which could be a non-radial g-mode.
  
  \end{itemize}
  
\item Various other secondary modes have been detected during a time interval of high density measurements, 
at 89, 129, 167, 185 and 234 d. 

\end{itemize}

Are radial and non-radial p- and g-modes simultaneously present in \object{V810~Cen} ? This is a quite fundamental
question because this supergiant would then be the first known star exhibiting such a pulsational behavior. In our
opinion, to answer this question, new hard and long-term work has to be done in the four following directions~:
{\it i)} New linear and non-linear non-adiabatic stability calculations have to be done, based on well-adapted supergiant
models; {\it ii)} High resolution (R$\simeq$40\,000) long-term spectroscopic monitoring must be organized in order to follow 
the line profile variations; {\it iii)} A long-term radial velocity monitoring would help to determine the parameters of the
orbit and of the components; {\it iv)} The photometric long-term monitoring must be continued in parallel to the
spectroscopic survey.

\begin{acknowledgements}
   We would like to express our warm thanks to all the observers at the Swiss 70~cm in La~Silla during the past 20 years. 
   This very long-term monitoring has been successful, thanks to their assiduity, maintained even with an air mass 
   larger than 3.5 ! We also thank our colleagues for the radial velocity measurements at the Danish 1.54~m in La~Silla.
   We are grateful to G\'erard Schaller for providing us with its non-adiabatic instability code. 
   We would like to thank T.~Aikawa, W.~Glatzel and A.~Gautschy for fruitful discussions about supergiants pulsations 
   as well as J.~Matthews and G.~Foster for their comments on the time serie analysis. The Centre de Données Astronomiques 
   de Strasbourg (CDS)  and NASA Data Archive and Distribution Service (NDADS) database are greatly acknowledged for providing us
   with \object{V810~Cen} spectra. This work has been partly supported by the Swiss National Science Foundation.
\end{acknowledgements}

\end{document}